\documentclass[12pt]{iopart}
\usepackage{graphicx}
\usepackage{subcaption}
\usepackage{float}
\usepackage[utf8]{inputenc} 
\usepackage{url}            
\usepackage{booktabs}       
\usepackage{nicefrac}       
\usepackage{microtype}      
\usepackage{array}
\usepackage[dvipsnames]{xcolor}

\expandafter\let\csname equation*\endcsname\relax

\expandafter\let\csname endequation*\endcsname\relax

\usepackage{amsmath}

\begin{document}
\title{Rare extinction events  in cyclic predator-prey  games}
\author{Shannon R Serrao}
\address{Center for Soft Matter and Biological Physics\\
  Department of Physics\\
  Virginia Tech\\
  Blacksburg, VA 24060 }
\ead{shann87@vt.edu}
\vspace{10pt}

\author{Tim Ritmeester and Hildegard Meyer-Ortmanns}
\address{
  Physics and Earth Sciences \\
  Jacobs University Bremen\\
  P.O. Box 750561, 28725 Bremen, Germany \\
}
\ead{t.ritmeester@jacobs-university.de}
\ead{h.ortmanns@jacobs-university.de}

\begin{indented}
\item[]March  2021
\end{indented}

\begin{abstract}
  In the May-Leonard model of three cyclically competing species, we analyze the statistics of rare events in which all three species go extinct due to strong but rare fluctuations. These  fluctuations are from the tails of the probability distribution of species concentrations. 
  They render a coexistence of three populations unstable even if the coexistence  is  stable in the deterministic limit.
  We determine the mean time to extinction by using a WKB-ansatz in the  master equation that represents the stochastic description of this model.  This way, the  calculation is reduced to a problem of classical mechanics and amounts to solving a Hamilton-Jacobi equation with  zero-energy Hamiltonian. We solve the corresponding Hamilton's equations of motion in six-dimensional phase space numerically by using the Iterative Action Minimization Method. This allows to project on the optimal path to extinction, starting from a parameter choice where the three-species coexistence-fixed point undergoes a Hopf bifurcation and becomes stable. Specifically for our system of three species, extinction events can be triggered along various paths to extinction, differing in their intermediate steps. We compare our analytical predictions with results from Gillespie simulations for two-species extinctions, complemented by an analytical calculation of the mean time to extinction in which the remaining third species goes extinct. From  Gillespie simulations we also analyze how the distributions of  times to extinction change upon varying the bifurcation parameter. Our results shed some light on the sensitivity of the times to extinction to system parameters. Even within the same model and the same dynamical regime, which allows a stable coexistence of species in the deterministic limit, the mean time to extinction depends on the distance from the bifurcation point in a way that contains the system size dependence in the exponent. It is challenging and worthwhile to quantify how rare the rare events of extinction are.
\end{abstract}

\section{Introduction}
Stochastic fluctuations play an important role in the survival or extinction of populations in ecological or biological contexts \cite{Bartlett1961StochasticPM,NisbetGurney,Allen2003AnIT}. Usually two types of fluctuations are distinguished: firstly external or environmental noise, and secondly, internal or demographic noise, induced by the discrete and probabilistic nature of the interactions between the constituents. Internal noise is therefore in principle unavoidable, its versatile effects cannot be ignored. In modelling approaches it leads to seemingly random deviations from the deterministic mean-field predictions of the system. An isolated finite population will eventually go extinct due to the nature of random fluctuations that drive changes in the populations (see references \cite{Assaf_2017,Ovaskainen2010StochasticMO} for an overview). Thus it is of much interest to understand  extinction as well as coexistence scenarios of populations being composed of several species which interact and compete with each other.  A focus should be on the timescales that govern these dynamics to quantify how rare the rare extinction events are and for how long a diversity of species can be maintained. One particularly relevant class of ecological models are the cyclic predator-prey models.

Cyclic predator-prey dynamics has been demonstrated in biological experiments \cite{ExperimentRPS}. These experiments show  either fixation of one species or coexistence of all species depending on the shape of the underlying substrate. On the theoretical side, ecological models \cite{May_Leonard,May73} with cyclic predator-prey dynamics are widely studied  due to the rich variety of phases they exhibit, or due the variety of spatio-temporal patterns which are observed in reaction-diffusion implementations of these models \cite{PhysRevE.87.032148,He2011,Reichenbach07mobilitypromotes,Serrao_2017,Labavic,Jiang}. Extinction events in a stochastic version of a cyclic Lotka-Volterra model have been studied in \cite{PhysRevE.74.051907}. The phase diagram of a stochastic version of the May-Leonard model, assigned to a spatial grid, was derived in \cite{PhysRevE.87.052710}. However, once we have to deal with rare extinction events, their nature, statistics, and relation to cyclically competing systems have not been addressed so far.

On a formal level, neither deterministic mean-field descriptions nor stochastic treatments on the level of Fokker-Planck equations  capture rare extinction events. The mean-field descriptions are suited to classify the different phases and to analyze the bifurcation diagram of the model. Fokker-Planck equations, as derived in a van Kampen-expansion \cite{vankampen} around the deterministic limit, determine the probabilities of observations  under the assumption that the stochastic fluctuations are small. The failure of these approaches is obvious from the fact that in stochastic experiments like Gillespie simulations \cite{Gillespie}, systems may escape from coexistence to extinction even in parameter ranges, in which mean-field equations predict a rather stable coexistence. A high stability is expected  far off from the bifurcation point (the point at which the coexistence gets unstable in the deterministic limit also without any fluctuations). Escapes from such a regime are indeed rare, because they require large fluctuations which are rather unlikely.

A suitable analytical framework to describe these rare events is the eikonal(-instanton) or WKB-approach (Wentzel-Kramers-Brioullin), usually familiar from quantum mechanics, but here applied to classical physics of rare events. For a review of this approach in the context of large deviations in stochastic populations we refer to reference \cite{Assaf_2017}. A seminal paper that initiated the use of the WKB-approach in the context of chemical kinetics is reference \cite{doi:10.1063/1.467139}, see also \cite{PhysRevLett.101.078101}. In our context, this approach provides the option to estimate  the mean time to extinction (MTE), since the calculation is mapped  to a problem of classical mechanics. Solving the classical Hamilton's  equations of motion yields the optimal path along which the system escapes. The classical action along this optimal path determines  the MTE, as we shall see.

In this paper, we consider extinction events in the May-Leonard model.
In contrast to a number of interesting previous works \cite{PhysRevE.78.060103,PhysRevE.70.041106,PhysRevE.79.041149,PhysRevE.81.021116} which considered populations of a single species, we have to deal with the competition between three species. Multiple species may in general lead to a proliferation of bifurcations in the phase diagram and disclose a number of alternative paths towards the same destination, which is full extinction of all species in our case. As we shall see, one can break up the  paths  into intermediate steps and, for example,  reduce the problem of finding the MTE of $n$ species to the MTE of $n-1$ species.  The overall MTE  of all species can be estimated by analyzing the complete set of paths to extinction.

We  study the  May-Leonard model for large stochastic fluctuations in a regime where (in the deterministic description) the coexistence-fixed point of three species is stable up to a Hopf bifurcation. At the Hopf bifurcation, the coexistence with highest diversity gets unstable, a regime that is not of interest in our context.

We will compare the calculated  times to extinction via the WKB-approach with average  times to extinction measured in Gillespie simulations of the same model. We shall see to what extent such a comparison  is possible. In more detail we analyze the  role  that is played by the distance from the Hopf bifurcation  for the MTE as well as for the distribution of  times to extinction. We also delineate the various paths to extinction and the role of multiplicity of the paths in obtaining the MTE.
\par
The paper is organized as follows. In section \ref{RES-Model} we present the model and the corresponding mean-field rate equations. In section \ref{RES:Fixedpoints}, we list the fixed points  of these  equations related to extinction, along with their stability and multiplicities. We next turn to the stochastic formulation in terms of a master equation and sketch the analytical approach that applies for rare large fluctuations in section \ref{RES:MEQ}. Rare events are those for which the time it takes them to happen is much larger than a typical time scale, here the relaxation time to approach the quasi-stationary distribution. It is based on the WKB-ansatz for the solution of the master equation, that is, the probability function to find n individuals of a certain species at a certain time. With this ansatz, the master equation transforms into a Hamilton-Jacobi equation. In section \ref{RES:OptimalPaths}, we identify the solutions of the associated Hamilton's equation of motion as optimal paths, minimizing the associated classical action that provides an estimate for the mean time to extinction. The optimal paths are trajectories in a six-dimensional phase space, their identification requires a refined numerical approach, the Iterative Action Minimization Method (IAMM) that is presented in section \ref{RES:NumericalAnalysis}. In section \ref{RES:Results} we present the results. From a plot of the trajectories in coordinate and momentum space as a function of time it is seen that the optimal paths to extinction have the shape of instantons. The results for the MTE are then compared with results for the MTE along another indirect optimal path, partially obtained  from Gillespie simulations, partially analytically. The comparison is subtle as both the WKB-ansatz and the Gillespie simulations face limitations of feasibility.  Furthermore, we show typical Gillespie trajectories and plot the histograms of times to extinction obtained from Gillespie simulations.  The histograms  reveals a dependence on the bifurcation parameter.  The results  indicate that one should take care on what is termed rare. We conclude in section \ref{RES:Conclusions}.

\section{The stochastic May-Leonard model}
\label{RES-Model}
The three-species  May-Leonard model \cite{May_Leonard,May73}, alternatively denoted as the (3,1)-model \cite{Labavic}, consists of three species interacting via three independent reactions, that is, a cyclically predating reaction with rate $\frac{\kappa}{V}$, a fitness reaction with growth rate $\rho$, and a deletion or intra-species competition reaction with rate $\frac{\gamma}{V}$ :
\begin{align}
\label{reactions}
X_\alpha + X_{\alpha+1} & \rightarrow  X_\alpha \textrm{ with rate }\   \kappa / V,  \nonumber \\
X_\alpha & \rightarrow  2 X_\alpha  \textrm{ with rate } \rho,  \nonumber \\
X_\alpha + X_{\alpha} & \rightarrow  X_\alpha \textrm{ with rate } \gamma / V.
\end{align}

Here, the index $\alpha = 1,2,3$ identifies the three different species and $X_{\alpha} \equiv X_{\alpha+3}$ with $X_\alpha$ representing an individual of species $\alpha$. $V$ denotes the volume, it parameterizes the system size in terms of the total number of individuals. In this work, we will simulate the set of reactions~(\ref{reactions}) via Gillespie simulations \cite{Gillespie}. The analytical treatment starts with a set of equations on the mean-field level in the sense that we ignore any spatial assignment of the corresponding concentrations of the individuals. Every species concentration can interact with every other  based on the rates of equation~$(\ref{reactions})$, which are set equal for all three species. In terms of the species concentrations $x_\alpha$ the mean-field rate equations  are written as
\begin{eqnarray}
\label{meanf}
    \frac{dx_1(t)}{dt}= \rho\ x_1(t) -\  \gamma\ x_1^2 (t) - \kappa\ x_1(t)\ x_3(t) \nonumber \\
    \frac{dx_2(t)}{dt}= \rho\ x_2(t) -\  \gamma\ x_2^2 (t) - \kappa\ x_2(t)\ x_1(t) \\
    \frac{dx_3(t)}{dt}= \rho\ x_3(t) -\  \gamma\ x_3^2 (t) - \kappa\ x_3(t)\ x_2(t). \nonumber
\end{eqnarray}

\section{Methods}
\subsection{Fixed point analysis of the system}
\label{RES:Fixedpoints}
The fixed points of the model are computed  from the mean-field equations by setting $\dot{x}_i = 0$ . The stability of these fixed points is determined by computing the eigenvalues of the linear stability matrix $ \mathcal{L}_{x_i^*}$ obtained from setting $\delta \dot{x}_i = \mathcal{L} \delta x_i$, where eigenvalues of $\mathcal{L}$ are evaluated at the fixed points $x_i^*$ of the system (for further details see reference \cite{Labavic}). The fixed points are summarized in  Table \ref{FP}.

\begin{table}[H]
\vline
    \begin{tabular}{c|c|c|c|}
    \hline
    Fixed points (FP) & Location(in units of V)      &   Eigenvalues (at FP)   \\ \hline
        FP$_1$ \text{(3 species extinction)}   &    (0,0,0)    &   $(\rho,\rho,\rho)$  \\ \hline
       FP$_2$ \text{(coexistence)}&    $(\frac{\rho}{\gamma + \kappa},\frac{\rho}{\gamma + \kappa},\frac{\rho}{\gamma + \kappa})$    &  $-\rho,\ \frac{-\rho(2\ \gamma - \kappa \ \pm i\ \sqrt{3} \kappa)}{2(\gamma + \kappa)}$  \\ \hline
        {FP$_{3-5}$ \text{(2 species extinction)}}  &   $(\frac{\rho}{\gamma},0,0)$    & $-\rho,\ \rho, \ \frac{\rho(\gamma-\kappa)}{\gamma}    $  \\ \hline
        {FP$_{6-8}$ \text{(1 species extinction)}} & $( \frac{\rho}{\gamma}, \frac{\rho(\gamma - \kappa )}{\gamma^2}, 0 )$ & $-\rho,\ \frac{-\rho(\gamma-\kappa)}{\gamma}, \ \frac{\rho(\gamma^2-\gamma\ \kappa +\ \kappa^2)}{\gamma^2}  $  \\
        \hline
    \end{tabular}
    \caption{Fixed points related to the extinction events. FP$_1$ is unstable independently of $\kappa$ and $\gamma$. FP$_2$ is stable for $\kappa < 2 \gamma$ and unstable for $\kappa > 2 \gamma$. FP$_{3-5}$ and FP$_{6-8}$ exchange the stability in one direction in a transcritical bifurcation at $\kappa=\gamma$.}
    \label{FP}
\end{table}
FP$_1$ denotes the fixed point where all species are extinct. It is unstable.
The second fixed point, the coexistence-fixed point FP$_2$, has one real eigenvalue that is always negative,
so it corresponds to a stable direction, and two complex eigenvalues whose real parts change sign at $\kappa=2\gamma$
 through a supercritical Hopf bifurcation. For $\kappa/\gamma< 2$ the fixed point is stable, otherwise it is
unstable in the directions corresponding to the pair of complex conjugate eigenvalues.
The other fixed points are always saddles, but the number of stable/unstable directions
depends on the ratio of $\kappa/\gamma$ and changes at $\kappa=\gamma$ through a transcritical bifurcation.
Fixed points FP$_{3-5}$ and FP$_{6-8}$ are two sets of three fixed points, each identical in nature, obtained by cyclically permuting their coordinates and eigenvalues. Fixed points FP$_{3-5}$, for which only one species is different from zero have always one
stable ($-\rho$-eigenvalue) and one unstable ($\rho$-eigenvalue) direction. The third direction
is stable for $\kappa<\gamma$, otherwise it is unstable.

Three fixed points FP$_{6-8}$, which correspond
to the survival of two species, also have always one stable (-$\rho$-eigenvalue) and
one unstable ($\rho\frac{\gamma^2-\gamma\kappa+\kappa^2}{\gamma^2}$-eigenvalue) direction.
The third direction changes sign at the
same point in phase space as in case of the previous fixed points, at $\gamma=\kappa$.
The direction corresponding to this eigenvalue is stable for
$\gamma>\kappa$, in contrast to the previous case. At $\gamma=\kappa$ the fixed points FP$_{3-5}$ and FP$_{6-8}$ collide and exchange the stability of one of the directions, so that for $\gamma>\kappa$ FP$_{3-5}$ has one stable and two unstable directions,
while FP$_{6-8}$ has two stable and one unstable direction. This happens in a transcritical bifurcation, at which the two sets  coalesce into one set of three fixed points. It should be noticed that for $\kappa>\gamma$, that is, above the transcritical bifurcation, FP$_{6-8}$ are physically irrelevant, since these fixed points correspond to negative concentrations of species.

Large fluctuations which drastically alter the state of the system are typically located in the tails of the probability distribution of the system's species concentrations, when the system is in a quasi-stationary state.
In particular, each of the possible rare  events or whole sequences of rare events encode a large fluctuation of the size $V$ as opposed to the smaller and frequent fluctuations of the size $V^{1/2}$ \cite{doi:10.1063/1.467139}.

\begin{figure}[h]
    \centering
    \includegraphics[width=10cm]{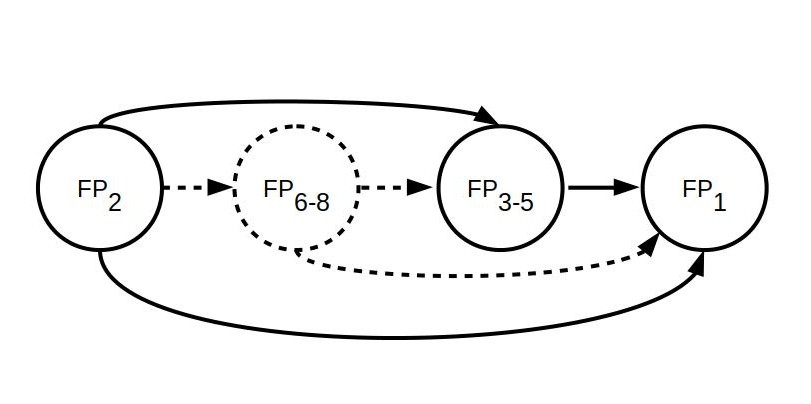}
    \caption{Possible paths to extinction. Solid lines indicate paths and fixed points (circles) that are always present, that is, independently  of $\kappa$ and $\gamma$. Dashed lines indicate paths and fixed points that correspond to a physically meaningful realization only for $\kappa <  \gamma$, that is, below the transcritical bifurcation point. }
    \label{fig:ExtPats}
\end{figure}

In this paper, we study the system in an initial state close to FP$_2$ in the stable parameter regime and compute a quantitative measure for the probability of extinction to a final state FP$_1$.
There are different options which the path to extinction can take.
A large fluctuation can either kick the system directly from FP$_2$ to FP$_1$, or  indirectly through other fixed points where only a proper subset of the population's species goes extinct. The various possible extinction routes are sketched in figure~(\ref{fig:ExtPats}). They depend on the rates $\kappa$ and $\gamma$. The MTE is then taken as a barometer for the  susceptibility to extinction, as is standard in the large deviation literature of ecological models.

\subsection{Master equation for the   May-Leonard model and WKB-ansatz}
\label{RES:MEQ}
The stochastic master equation for a general set of reactions changing population concentrations ${ n_\alpha}$ by $r_\alpha$ per time step is given by
\begin{align}\label{GMEQ}
    \frac{\partial P(n_{\alpha};t)}{\partial t} = \sum_{\alpha} \biggl\{ W(n_{\alpha}\ -\ r_\alpha ; r_\alpha)\  P(n_{\alpha} -\ r_\alpha ;t) - W(n_{\alpha}\  ; r_\alpha)\  P(n_{\alpha} ;t) \biggr\},
\end{align}
where $P(n_{\alpha};t)$ denotes the probability of finding $n_\alpha$ individuals of species $\alpha$ at time $t$ and $W(n_\alpha; r_\alpha)$ is the rate for the given reaction.

The master equation for the  May-Leonard model is given as
\begin{align} \label{MEQ}
 \frac{\partial P(n_{\alpha};t)}{\partial t} & =  \sum_{\alpha=1,2,3} \biggl\{ \frac{\kappa}{V}\ \bigl\{ n_{\alpha}\ (n_{\alpha +2}+1)P(n_{\alpha},n_{\alpha +2}+1;t) -n_{\alpha}n_{\alpha +2} P(n_{\alpha},n_{\alpha +2};t) \bigr\} \nonumber \\
&+  \rho\ \bigl\{
   (n_{\alpha}-1)P(n_{\alpha}-1;t) - n_{\alpha}\ P(n_{\alpha};t) \bigr\} \\ \nonumber
&+ \frac{\gamma}{V} \bigl\{n_{\alpha}\ (n_{\alpha}+1)\ P(n_{\alpha}+1;t) - n_{\alpha} \ n_{\alpha}\ P(n_{\alpha};t) \bigr\} \biggr\}.  
\end{align}
In the notation we have spelt out two arguments $n$ in $P$ (such as in $P(n_{\alpha},n_{\alpha +2};t)$) to indicate which concentrations are involved in the respective reaction.
For $n_\alpha>0$ and times $t$ much larger than the initial relaxation time $t_{rel}$ towards a quasi-stationary distribution, determined by the deterministic rate equation, the overall time dependence for the long-time decay is expected according to
\begin{equation}
\label{MTE_comp}
P(n_\alpha>0, t) = P_{st}(n_\alpha)\cdot e^{-t/\tau},
\end{equation}
where $1/\tau$ with $\tau>>t_{rel}$ is the lowest positive eigenvalue of the master equation and $\tau$ the decay time. $P_{st}$ denotes the quasi-stationary distribution. For metastable populations, the decay time is a good approximation to the mean time to extinction (MTE) that we calculate later. It sets the typical time scale during which the distribution remains quasi-stationary.
Following the work of Dykman \textit{et al.} \cite{doi:10.1063/1.467139}, we use now the WKB-ansatz of the form
\begin{equation}
 P_{st}(n_\alpha) \propto e^{-s_{st}(n_\alpha)V}
 \label{WKBeq}
\end{equation}
with scaled stationary action $s_{st}(n_\alpha)=S_{st}(n_\alpha)/V$ as a function of all species concentrations, which will be evaluated along an optimal path in the phase space of concentrations $n_{\alpha}$ and auxiliary momenta $p_\alpha$, to be defined next. Usual, small fluctuations are of the order $\sqrt{V}$. However, this ansatz should project on the tails of the distribution that correspond to large fluctuations in the number of individuals of the population, they are of order $V$ and may lead to extinction events which are rare.  Expanding the action to $\mathcal{O}(1)=\mathcal{O}(V^0)$ in $1/V$, we have $S_{st}(n_\alpha + r_\alpha) \cong S_{st}(n_\alpha) + \sum_\alpha r_\alpha \frac{\partial}{\partial n_\alpha } S_{st}(n_\alpha) $.
We use scaled variables of species concentrations   $x_\alpha=n_\alpha/V$, and  define conjugate momenta as $p_\alpha \equiv \frac{\partial s_{st}}{\partial x_\alpha}$. Plugging in the ansatz for the action in terms of the scaled variables into equation~(\ref{MEQ}) and neglecting  terms of $O(1/V)$, we obtain the following expression:
\begin{eqnarray} \label{MEQ:QuasistationaryApproximation}
 \sum_{\alpha=1,2,3} &\biggl\{&\frac{\kappa}{V} \bigl\{V^{2} x_{\alpha} (x_{\alpha +2})e^{-s_{st}({x_\alpha})V -\frac{\partial}{\partial x_{\alpha+2} } s_{st}({x_\alpha})V} -V^{2} x_{\alpha}x_{\alpha +2} e^{-s_{st}({x_\alpha})V} \bigr\}  \nonumber \\
&+& \rho\ V\ \bigl\{(x_{\alpha})e^{-s_{st}(\{x_\alpha\})V+\frac{\partial}{\partial x_\alpha } s_{st}(\{x_\alpha\})V} - x_{\alpha}\ e^{-s_{st}(\{x_\alpha\})V} \bigr\}   \\
&+&\frac{\gamma}{V} \bigl\{(V^{2}\ x_{\alpha}^{2}\ e^{-s_{st}(\{x_\alpha\})V -\frac{\partial}{\partial x_\alpha } s_{st}(\{x_\alpha\})V} - V^{2}\ x_{\alpha}^{2}\ e^{-s_{st}(\{x_\alpha\})V} \bigr\} \biggr\} \approx 0.   \nonumber
\end{eqnarray}
This amounts to a Hamilton-Jacobi equation $H(x_\alpha,\frac{\partial s_{st}}{\partial x_\alpha})=0$ with Hamiltonian $H(x_\alpha, p_\alpha)$  which yields for the May-Leonard system:
\begin{eqnarray}
\label{Hamiltonian}
H(x,p)=\sum_{\alpha=1}^3 \Bigg[\rho x_\alpha(e^{p_\alpha} -1) + (\gamma\  x_{\alpha}^2 +\kappa\  x_\alpha\ x_{\alpha+2})\cdot(e^{-p_\alpha}-1)\Bigg]=0.
\end{eqnarray}

Specific solutions of the associated Hamilton's equations of motion for $x_\alpha, p_\alpha$ will include the optimal path to extinction.

\subsection{Optimal paths for extinction trajectories of the May-Leonard model}
\label{RES:OptimalPaths}
The trajectories of the system can be computed by solving the Hamilton's equations of motion of the system as computed from equation~(\ref{Hamiltonian}):
\begin{align}
\label{HE}
\dot{x}_\alpha & = \frac{\partial H}{\partial p_\alpha}  =  \rho\ x_\alpha\ e^{p_\alpha} - \gamma\ x_\alpha^2\ e^{-p_\alpha}\ -\ \kappa\ x_\alpha\ x_{\alpha+2}\ e^{-p_\alpha}  \nonumber \\
-\dot{p}_\alpha & = \frac{\partial H}{\partial x_\alpha}  =   \rho\ (e^{p_\alpha}-1) + (e^{-p_\alpha}\ -1)(2\ \gamma\ x_\alpha\ +\ \kappa\ x_{\alpha+2})   +(e^{-p_{\alpha+1}}\ -1)(\kappa\ x_{\alpha+1}). \\ \nonumber
\end{align}
The solutions $(x_{\alpha}(t), p_{\alpha}(t))$ define paths that are taken by the system and minimize the action for given initial and final coordinates and momenta. Here we start from the coexistence-fixed point $FP_2$. As long as the momenta $p_\alpha$ are not fixed, every member of this set of paths has a different likelihood. For the destination at FP$_1$ the path should end at zero coordinates and momenta. The optimal trajectory will be inserted into the scaled action $s_{st}(x_\alpha)$  associated with the Hamiltonian:
\begin{equation}
    s_{st}(x_\alpha) \equiv \int_{t_0}^t \it{dt'} \textit{L}(x_\alpha, \dot{x}_\alpha) = \int_{t_0}^t \it{dt'} \sum_{\alpha} p_{\alpha} \dot{x}_{\alpha} - H(x_\alpha, p_\alpha),
\end{equation}
which reduces to
\begin{equation}
 s_{st}(x_\alpha) =\int_{t_0}^t \it{dt'} \sum_{\alpha} p_{\alpha}\dot{x}_\alpha =\int_{x_{st}}^x  \sum_{\alpha} p_{\alpha}\ dx'_{\alpha},
 \label{act}
\end{equation}
since $H(x_\alpha, p_\alpha) = 0 $. Here $x_{st}$ denote the normalized concentrations at the stable fixed point FP$_2$, chosen as the initial point of the path.
The optimal path between initial and final point in phase space is the most likely path to extinction, it minimizes the action equation~(\ref{act}). It is computed numerically for given boundary conditions.

In principle, one may think of calculating the action analytically by splitting the variables in slow and fast degrees of freedom  in the vicinity of the bifurcation point in analogy to reference~\cite{doi:10.1063/1.467139}; there one has a single slow mode over which the action is integrable plus a quadratic function corresponding to the fast modes, where the latter term leads to a Gaussian shape in the associated quasi-stationary probability distribution in phase space.  However, the May-Leonard system has a pair of slow degrees of freedom with complex eigenvalues, the real part of which vanishes at the Hopf bifurcation.   Our system with two slow modes is not integrable, thus we cannot analytically establish the power-law scaling on the slow manifold  near the bifurcation.
\par
Keeping the  fixed points  as the initial and final destinations (from coexistence at FP$_{2}$ to total extinction at FP$_{1}$), the process to extinction can still take several routes (see figure~\ref{fig:ExtPats}). In general, what is accessible via Gillespie simulations are paths which are realized after a time that does not increase exponentially with the system size, unless the system size is multiplied by a small factor (in our case the scaled action from FP$_2$ to FP$_{3-5}$). In this sense only the fastest routes are  seen in Gillespie simulations  of the model.  Here this will be the path from FP$_2$ to FP$_{3-5}$. Within the WKB-approach, we solve equations (\ref{HE}) via numerical integration from FP$_{2}$ directly to FP$_{1}$ once  we  specify a stable initial guess of the path for the numerical solver. For the direct path, we solve the Hamilton's  equations~(\ref{HE}) using the quasi-Newton based IAMM approach, based on an initial guess of the optimal path. The method relies upon a good initial guess. For extinction paths which involve intermediate fixed points we did not succeed in finding a viable initial guess. Therefore, the indirect path from FP$_{2}$ to FP$_{1}$ via FP$_{3-5}$ or FP$_{6-8}$ was not fully analytically feasible due to the part from FP$_{2}$ to FP$_{3-5}$ or FP$_{6-8}$. The reason is that initially the path from FP$_2$ in the vicinity of the bifurcation point shows increasing oscillations. They result from the fact that the real part of a pair of complex eigenvalues goes to zero, as seen in a linear stability analysis around the fixed point. This requires a fine discretization of the path, but the numerical error remains highly sensitive to small changes of the path, so that a convergence to the optimal path failed.
However, just the path from  FP$_{2}$ to FP$_{3-5}$ is numerically accessible via Gillespie  simulations  for the relevant parameter range.

Vice versa, for a direct path to full extinction (to FP$_{1}$) or for part of the route from  FP$_{3-5}$ to  FP$_{1}$,  Gillespie simulations are not accessible, as the simulation time of the Gillespie algorithm blows up: The scaled action along the respective optimal paths turns out to be of the order of $1$ (see figure 4 below) and the volume-term is large for sufficiently many individuals (factors like $\sim e^{1000}$ occur). On the other hand, the system at the stage of FP$_{3-5}$ reduces effectively to a single-species system that can go extinct, and here the MTE is again analytically accessible.

The estimation of extinction times from FP$_{2}$ to FP$_{1}$ via FP$_{3-5}$ is therefore done in two steps: First the Gillespie simulations are employed to compute the time to single-species extinction from FP$_{2}$ to FP$_{3-5}$, followed by an analytical estimation of the extinction time in an effectively one-dimensional model  from  FP$_{3-5}$ to FP$_1$, using again the WKB-ansatz \cite{doi:10.1063/1.467139}.\par

Quasi-Newton methods like the Iterative Action Minimization Method \cite{LINDLEY201322}) are known from solving non-linear systems of equations. They  are employed also here to solve equation~(\ref{HE}). The trajectories are computed starting from the coexistence-fixed point  FP$_{2}$ and  zero momentum $({\bf{x^*}},{\bf{0}})$, since zero momentum corresponds to the  point in phase space that is shared with the deterministic limit.  The final destination is the extinction point FP$_{1}$ with zero momentum, passing close to FP$_1$ at large negative momentum, whose exact value depends on the accuracy of the algorithm, see also the remarks below. The higher the accuracy, the more negative this value is in accordance with $\frac{dp}{dx}\rightarrow -\infty$ as FP$1$ is approached. 
\par
Once we know the value of the scaled action $s_{st}$ of (\ref{act}) along the optimal path, we  compute the MTE from coexistence to extinction as follows. If $\tau$ in equation~(\ref{MTE_comp}) is identified with the MTE, we integrate equation~(\ref{MTE_comp}) (using that the probability integrates to one) to obtain \cite{PhysRevE.81.021116}:
\begin{equation}
     \text{MTE} \propto e^{s_{st}\ V }.
     \label{MTE}
 \end{equation}
This leads to a quantitative measure for the MTE  if the extinction is a fluctuation that is based on a rare event.

\subsection{Numerical analysis of the Hamiltonian dynamics and Gillespie simulations}
\label{RES:NumericalAnalysis}
\subsubsection{Optimal trajectories using IAMM from FP$_2$ to FP$_1$ along the direct path}
\label{RES:OptimalTrajIAMM}
 We follow the numerical scheme coined Iterative Action Minimization Method (IAMM) as outlined in \cite{LINDLEY201322,PhysRevE.97.012308} to compute the extinction paths numerically. This technique provides a direct and explicit iterative scheme of computing the optimal path in higher dimensions. We summarize the involved steps:
\begin{itemize}
    \item The optimal path lies on the zero energy surface defined by the Hamilton's equations~(\ref{Hamiltonian}) from the coexistence point FP$_{2}$ to FP$_{1}$.
    Let the path start at FP$_{2}$ at $t =-\infty $ and end at FP$_{1}$ at $t =\infty $. We assume that the very transition occurs over a small time interval relative to the total time and therefore map the transition from FP$_{2}$ to FP$_{1}$ to an interval of size $2\ T_\epsilon$:
    \begin{align*}{}
        FP_{2}  &  \textrm{ at } (-\infty, - T_\epsilon] \nonumber \\
        FP_{2} \rightarrow FP_{1} & \textrm{ at }  (- T_\epsilon, + T_\epsilon) \nonumber \\
        FP_{1}  & \textrm{ at } [ T_\epsilon, \infty) . \nonumber
    \end{align*}
     Next we map the interval $[- T_\epsilon, + T_\epsilon]$ onto $[0,1]$, with the transformation $t=2 T_{\epsilon} h- T_{\epsilon}$, and simulate the system from $h\in [0,1]$ (since we are interested in the action so that the actual rescaling of time is irrelevant.)
    \item Given a  value of the bifurcation parameter $\kappa$, for example at the bifurcation point $\kappa=2$ with $\gamma=1$. The Hamilton's  equations~(\ref{HE}) are solved for the above interval via uniformly dividing the interval $[0,1]$ into $N$ steps. We used an adapted version of the Matlab code provided in \cite{PhysRevE.97.012308}: For the discretized system of equations on a mesh,  a central difference scheme is used employing the eighth-order approximations  for better accuracy \cite{Finitediff}. This scheme converts the set of differential equations to algebraic equations, which are solved using the ``Trust-region-dogleg'' method, as implemented by Matlab's `fsolve' module.  
        As an initial guess for the path (not only suited at the bifurcation point, but also beyond) we adapt the suggestion of \cite{LINDLEY201322}) according to
\begin{equation}
x_\alpha(h)=\frac{x^*_\alpha}{1+\exp{[40(h-0.5)]}}, \qquad p_\alpha=0.
\end{equation}
The optimal path is then found by this iterative procedure for one $\kappa$-value.
\item The previous steps are then repeated for different $\kappa$-values. Here one may choose the optimal path for the previous $\kappa$-value  as initial guess for the optimal path at the next  $\kappa$. This procedure constitutes the IAMM: It computes the optimal path $x(t)$ and $p(t)$ for parameter values of the entire range of the bifurcation parameter.
\end{itemize}
    Since independently of the values of the bifurcation parameter, all paths should lie asymptotically close to the zero-energy surface and correspond to the most unstable directions along which the system most easily escapes, an appropriate initial guess is on the one hand essential, but on the other hand challenging  to guess because of the high sensitivity of the optimal path to deviations from it. A derivation of an analytical form of the initial guess according to reference~\cite{PhysRevE.97.012308} and references therein does not seem to be applicable to our case because of the distance between the fixed points and their local neighborhoods.

\subsubsection{From  FP$_2$ to FP$_{1}$ along the indirect path via FP$_{3-5}$}\label{RES:OptimalTrajGillespie}
We cannot yet compare our results from the WKB-ansatz to the standard Gillespie simulations which capture the fully stochastic nature of these kinetic reactions. The WKB-ansatz yields the trajectories and hence the action from the coexistence-fixed point FP$_2$ at populations $n_1=n_2=n_3=(\frac{\rho}{\gamma  +\kappa})\ V$ directly to the three-species extinction. In view of a comparison of the orders of magnitude, we have next to determine the action  for the MTE from FP$_2$  to  FP$_{3-5}$ from the results of the Gillespie simulations, complemented by an analytical calculation of the MTE from
FP$_{3-5}$ to FP$_{1}$. The results for the direct and indirect path need not agree, but should be compared.

\begin{figure}[h]
    \centering
    \includegraphics[height=8cm,width=13cm]{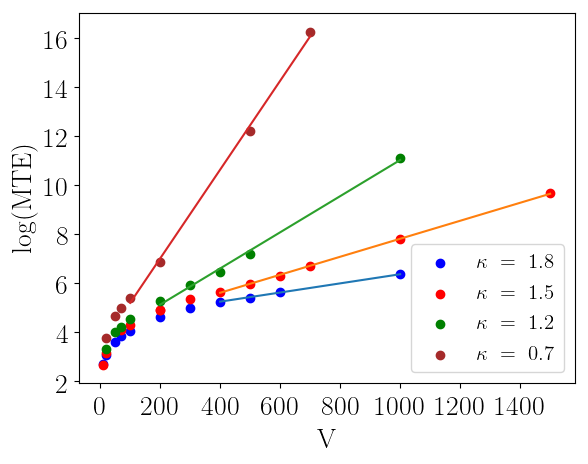}
    \caption{Computing the action $s_{2D}$ from the MTE based on Gillespie simulations, for various values of $\kappa$, for $\gamma=1$ and $\rho=0.1$. The action is read off from the slope of the line fitted to the asymptotic part of the curves, neglecting points for small $V$ where sublinear contributions to the MTE become significant.}
    \label{fig:MTEGillespie}
\end{figure}

\paragraph{From  FP$_2$ to FP$_{3-5}$ via Gillespie simulations}
We apply the Gillespie algorithm to the set of reactions of equation~(\ref{reactions})  and measure
the time taken to reach single-species coexistence, that is, two-species extinction. We always start from the position of FP$_2$. A two-species extinction event is registered if two out of the three populations go  exactly to zero. This signals the arrival at one of the three fixed points FP$_{3-5}$, characterized by a single species surviving. Since all three predation rates are equal, the two out of three species that go extinct in a given simulation can be any of the three options. Using ensembles of $10,000$ runs we average over the time to extinction to obtain the MTE from FP$_2$ to FP$_{3-5}$. This estimate is then converted into an estimate  of the part $s_{2D}$ of the total scaled action $s_{st}$ (according to equation~(\ref{MTE})), by computing the slope at the asymptotic values of the logarithm of the MTE, see figure~\ref{fig:MTEGillespie}. The subscript $2D$ refers to the two  species that go extinct.

Since the Gillespie simulations of the system evolving from FP$_{3-5}$ to FP$_{1}$ become prohibitively expensive (as the action there is of $O(1)$ (see below)),  we next compute the MTE for this path  analytically via the action $s_{1D}$, again  for large values of the volume $V$. This is easily done, as the dynamics becomes one-dimensional. The reason is that there is no process in the May-Leonard model that will retrieve a population once it got extinct, thus the irreversibility of extinction guarantees the reduction to one dimension along the route from FP$_{3-5}$ to FP$_{1}$. This is also seen from Hamilton's equations of motion that there exists an invariant subspace defined by the two extinct species, in which the surviving species resides.

\subsubsection{From FP$_{3-5}$ to FP$_{1}$ via a derivation of the action for a one-species extinction problem}
When two species get extinct at FP$_{3-5}$, the system effectively reduces to a one-dimensional problem in one of the three species, $\alpha \in \{1,2,3 \}$, with reactions $X \xrightarrow{\rho} 2X $ and the reverse reaction  $ 2X \xrightarrow{\gamma} X $. The Hamiltonian for this system is given as
\begin{equation}
    \label{1DH}
    H(x,p) = \rho\ x (e^{p}-1) + \gamma\  x^2\ (e^{-p}-1) =0.
\end{equation}
This equation is solved for $p(x)$ and the scaled action is computed as
\begin{equation}
    \label{1Ds}
    s_{1D} = \int_{FP_{3-5}}^{FP_{1}} p\ dx = \int_{\frac{\rho}{\gamma}}^{0^+} log( \frac{\gamma}{\rho\ }\ x)dx.
\end{equation}
Integrating the action from the single-species extinction FP$_{3-5}$($\frac{\rho}{\gamma}$) to FP$_{1}$($0^+$), we have  $s_{1D}= \frac{\rho}{\gamma}$. Therefore, the corresponding MTE$_{1D} \propto e^{\frac{\rho}{\gamma}V}$. Here it should be noticed that $\frac{\rho}{\gamma}\sim O(1)$, which explains why it is hopeless to wait for extinction events in the Gillespie simulations if the volume is of the $O(1000)$ or larger. Thus the total MTE for all three species along the indirect path is given as
\begin{equation}
    \text{MTE} \propto e^{(s_{1D} +  s_{2D})\ V}.
    \label{MTE_total}
\end{equation}
Below we will compare the scaled action $s_{st}$ with the values of $(s_{1D} +  s_{2D})$ from equation~(\ref{MTE_total}) along direct and indirect paths.

\section{Results}
\label{RES:Results}

\subsection{Trajectories in six-dimensional phase space}

\begin{figure}
\begin{subfigure}{1\textwidth}
   \includegraphics[height=5cm,width=15cm]{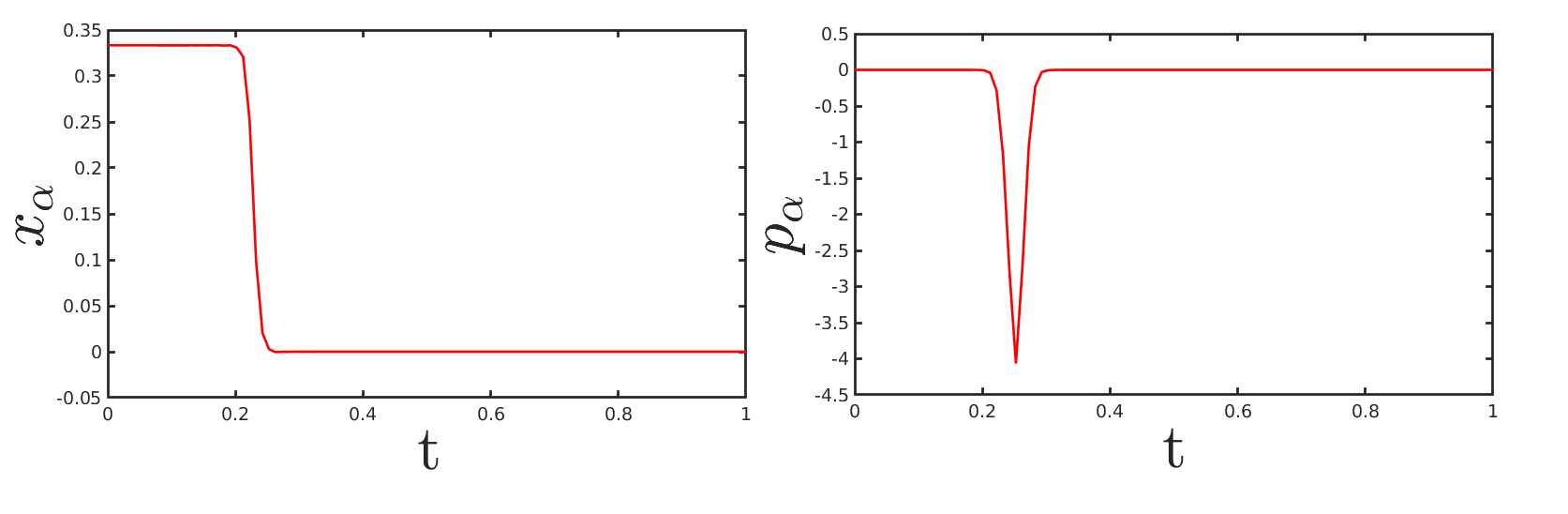}
\caption{}
  \label{fig:traj-first}
\end{subfigure}
\begin{subfigure}{1\textwidth}
  \centering
  \includegraphics[height=5cm,width=15cm]{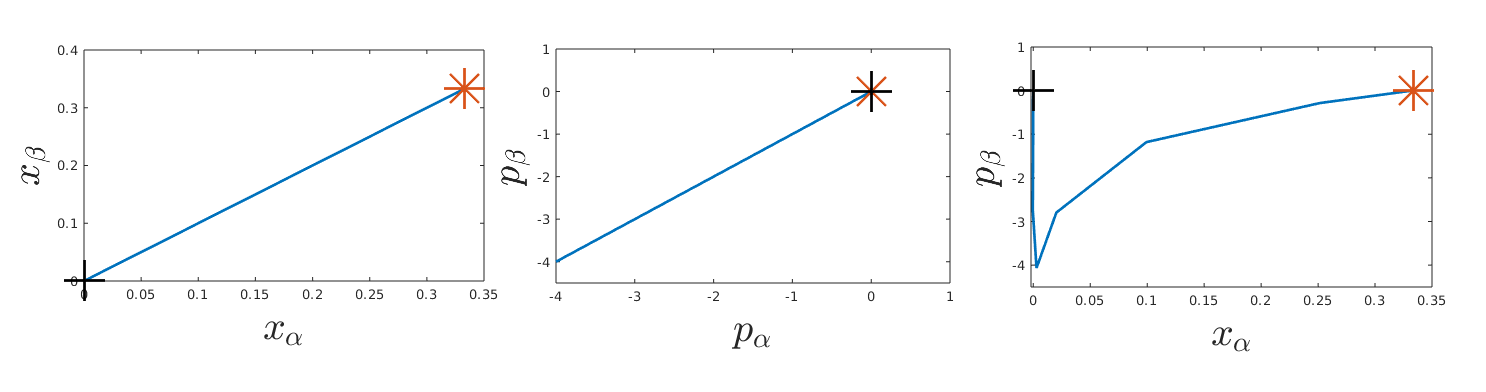}
\caption{}
  \label{fig:traj-second}
\end{subfigure}
\caption{Optimal paths  of the large fluctuations toward extinction in the May-Leonard model. Figure~(\ref{fig:traj-first}) (left) shows an instanton trajectory of species concentrations $n_\alpha$ with $\alpha=1,2,3$ along with conjugate momenta (right panel) from  FP$_2$ to FP$_{1}$ for $\kappa = 2$, $\gamma = 1$, $\rho=1$, mapped to the time interval $t\equiv h \in[0,1]$. The size of the peak to negative values in the momentum depends on the accuracy of the algorithm, its position within this time interval is arbitrary. Figures~(\ref{fig:traj-second}) depict phase space trajectories according to the WKB-ansatz from FP$_{2}$ (red `$\ast$') to FP$_{1}$ (black `+') from the coexistence point FP$_2$ to FP$_{1}$ for $\kappa = 2$, $\gamma = 1$, $\rho=1$, and $\alpha,\beta\in\{1,2,3\}$.}
\label{fig:traject}
\end{figure}

In general, optimal trajectories computed via the WBK-ansatz show instanton-like trajectories in  phase space, this is also seen here  in figure~(\ref{fig:traj-first}), obtained from the integration of the Hamilton's equations of motion. Typically the extinction of all three species is associated with a large fluctuation in the system which is inferred from the large negative momentum spike in figure~(\ref{fig:traj-first}). The negative value $-4$ in the figure depends on the applied accuracy, here chosen as $10^{-5}$. For machine accuracy it becomes about $-7$ and diverges when the kink in the position corresponds to an actual jump. The divergence is also seen from equation~(\ref{1Ds}), when $x$ approaches zero in the argument of $\log$. This rare event causes all three species to fall to zero.  After the negative peak in the momentum, the system  evolves in the opposite direction of the deterministic path (recall that in the deterministic limit FP$_1$ is unstable), so it moves along $x=0$ within the numerical accuracy to FP$_1$, where $x=0$, $p=0$, cf. figure~(\ref{fig:traj-second}) most right. The figure shows a typical projection that holds for all species $\alpha,\beta\in\{1,2,3\}$. It should be mentioned that the results shown in this figure, which were obtained directly at the bifurcation point, are very similar in the immediate vicinity of the bifurcation point.

\subsection{Quantitative comparison of the MTEs determined via  the scaled actions}
\label{RES:ComparisonAction}
\begin{figure}[h]
    \centering
    \includegraphics[height=8cm,width=15cm]{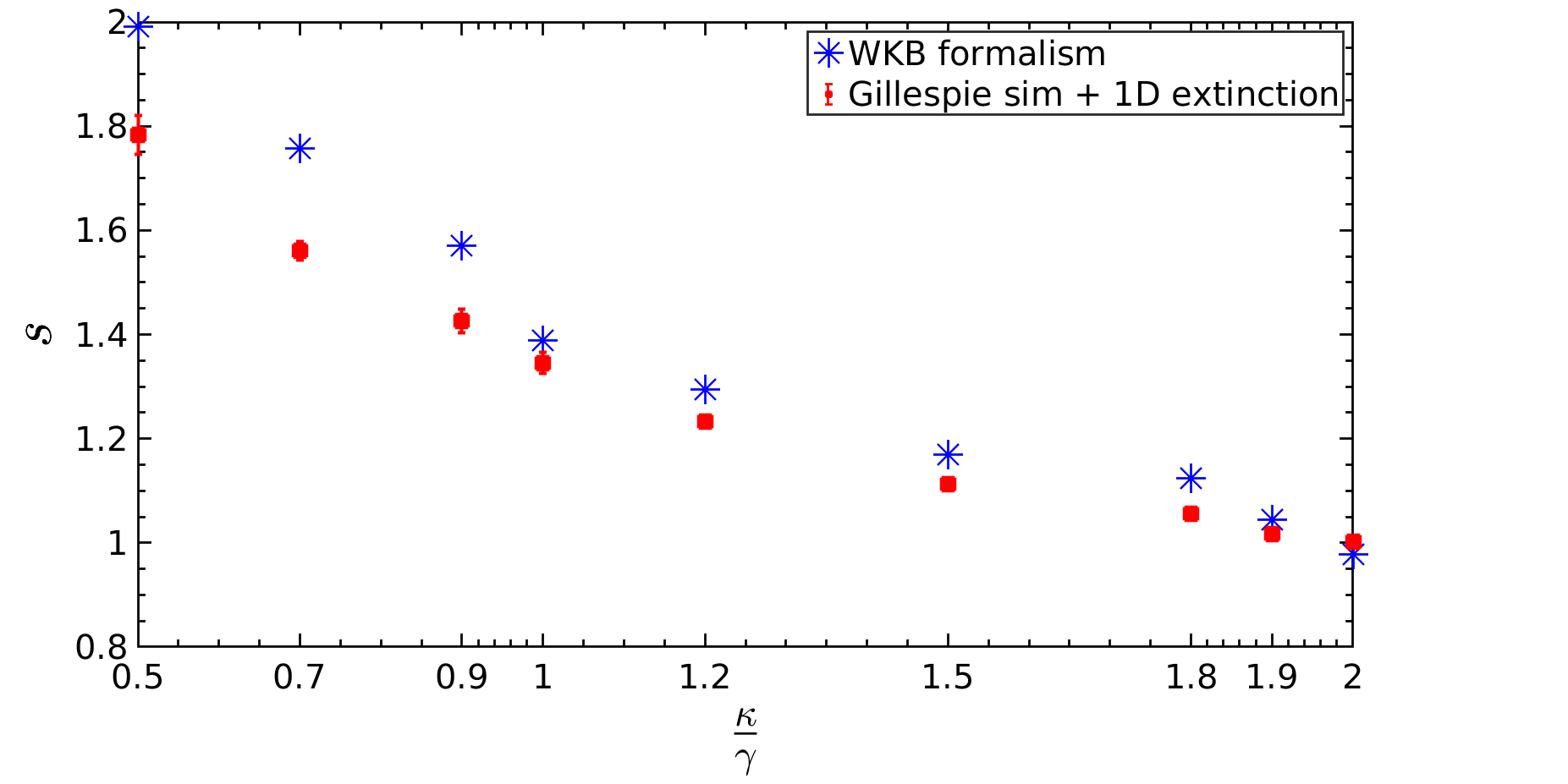}
    \caption{Comparison of the scaled action along the optimal path from the WKB-ansatz and the combined result of the Gillespie simulation  with the solution of the one-dimensional problem,
    as a function of $\frac{\kappa}{\gamma}$ which parameterizes the distance from the Hopf bifurcation at $\kappa=2\gamma$. Error bars for the Gillespie simulations, resulting from the fits of figure~\ref{fig:MTEGillespie}, are small for $\kappa/\gamma > 1.0$ ($\approx 10^{-4}$) and hardly visible. $\rho=\gamma=1$. }
    \label{Action_comparison}
\end{figure}

We want to quantitatively compare the MTE obtained on the one side from the action $s_{st}$ for the direct path from FP$_2$ to FP$_1$ via the WKB-method alone, and on the other side from the combined action $s_{1D} +  s_{2D}$ as in equation~(\ref{MTE_total}) for the indirect path from FP$_2$ via FP$_{3-5}$ to FP$_1$. Since the actions provide an asymptotic estimate of the corresponding MTEs, we use directly the actions as means for comparing our analytical and simulation results.
We see very similar values in figure~(\ref{Action_comparison}). They actually  match  in the vicinity of the bifurcation point, although the escape along the direct and indirect paths may take different times, in principle. For smaller values of $\kappa/\gamma$, in particular below the transcritical bifurcation at $\kappa/\gamma=1$, the MTE according to the WKB-method lies  above the results from the partially numerically obtained results. Here the reason is obvious: As soon as the numerically registered extinction events include alternative existing paths toward full extinction via the fixed-points FP$_{6-8}$,  the WKB-method  overestimates the MTE as long as it does not include these paths. For small values of $\kappa/\gamma$, the contributions of $s_{1D}$ and $s_{2D}$ are of the same order of magnitude. Vice versa, Gillespie simulations may in principle underestimate the MTE if this time scale should  refer to an event that is caused by a really large and rare fluctuation towards extinction.

Near and at the Hopf bifurcation, the scaled action  takes the smallest values. This is expected from the fact that close to the Hopf bifurcation an escape via the two-dimensional slow manifold is facilitated. Near the Hopf bifurcation, a pair of complex conjugate eigenvalues has a  real part close to zero, the corresponding two eigenmodes are  slow modes of the system. Along these eigen-directions, large fluctuations are more likely. This reduces the MTE.  As we move away from the bifurcation point, an escape from the coexistence-fixed point FP$_{2}$ becomes harder as it requires larger fluctuations to induce extinction which are here more rare.

From  figure~(\ref{Action_comparison}) and $s_{1D}=1$, we infer that near the Hopf bifurcation, the time to three-species extinction is dominated by the single-species extinction. At the Hopf bifurcation, FP$_2$ is a center and the system easily goes to two-species extinction. Thus, the rare fluctuations that lead to complete extinctions  and give the main contribution to the total scaled action from FP$_2$ via FP$_{3-5}$ to FP$_1$ result mainly
from the transitions FP$_{3-5}$ to FP$_{1}$. The times to escape via the direct and indirect paths need not agree, but  their values do agree at the bifurcation point. From the instanton in figure~(\ref{fig:traj-first}) (left) we see that nevertheless the paths do not coincide, as the direct one reaches FP$_1$ without passing one of the FP$_{3-5}$ fixed points. The MTE for the single-species problem is $\mathcal{O}(e^{s_{1D}V})$ with $s_{1D}=\rho/\gamma=1$ for our choice of parameters $(\rho=1, \gamma=1)$. In the regions away from the bifurcation,  the scaled actions differ between $2.0$ (WKB) and $1.8$ (Gillespie) at $\kappa=0.5$, respectively, and $1$ (WKB) and $1.05$ (Gillespie) at $\kappa/\gamma=2$, so that in the intermediate regime  the path from FP$_{2}$ to FP$_{3-5}$ does contribute to the MTE.\\

Although the results from Gillespie simulations refer only to the restricted path  from FP$_2$  to FP$_{3-5}$, the dependence of the shape of the distribution on the bifurcation parameter  sheds some light on the general need for further specifying the nature  of rare fluctuations. Within the same model and even within the same dynamical regime of  stable coexistence in the deterministic limit, the dependence of rare fluctuations on the model parameters  has an impact on the MTEs. As the scaled action in figure~\ref{Action_comparison} changes roughly by a factor of $2$ as a function of $\kappa/\gamma$, the MTE increases  by about a factor of $e^V=e^{2V}/e^V$ in the considered range of parameters in figure~\ref{Action_comparison}.

\subsection{Trajectories of Gillespie simulations}
It is instructive to pursue the fluctuations in Gillespie trajectories in which fixed points of the deterministic limit become clouds of data points of the stochastic trajectories in three-dimensional configuration space
\begin{figure}[!htb]
\minipage{0.33\textwidth}
  \includegraphics[width=\linewidth]{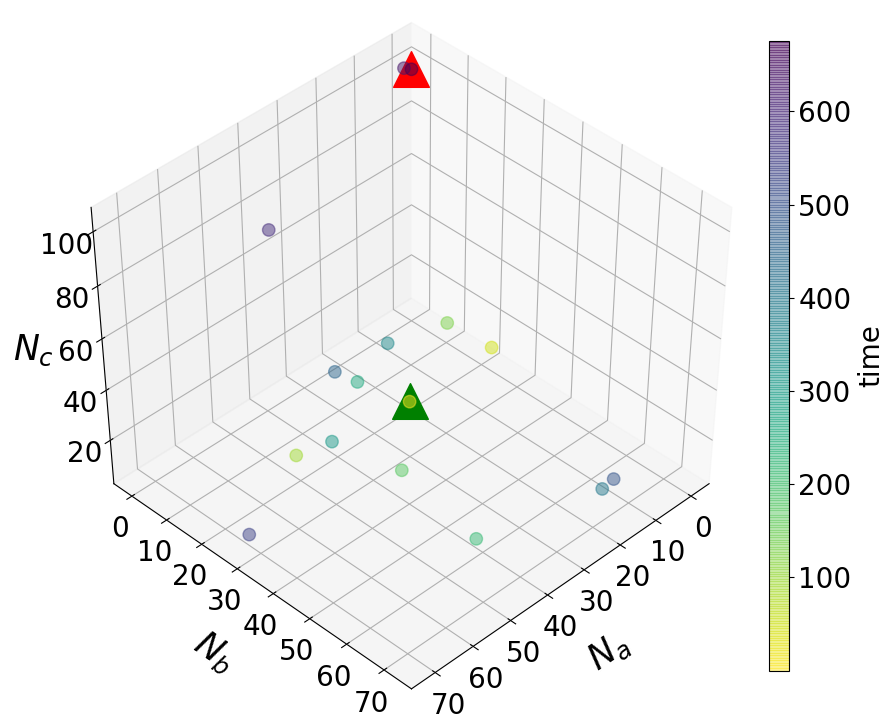}
\endminipage\hfill
\minipage{0.33\textwidth}
  \includegraphics[width=\linewidth]{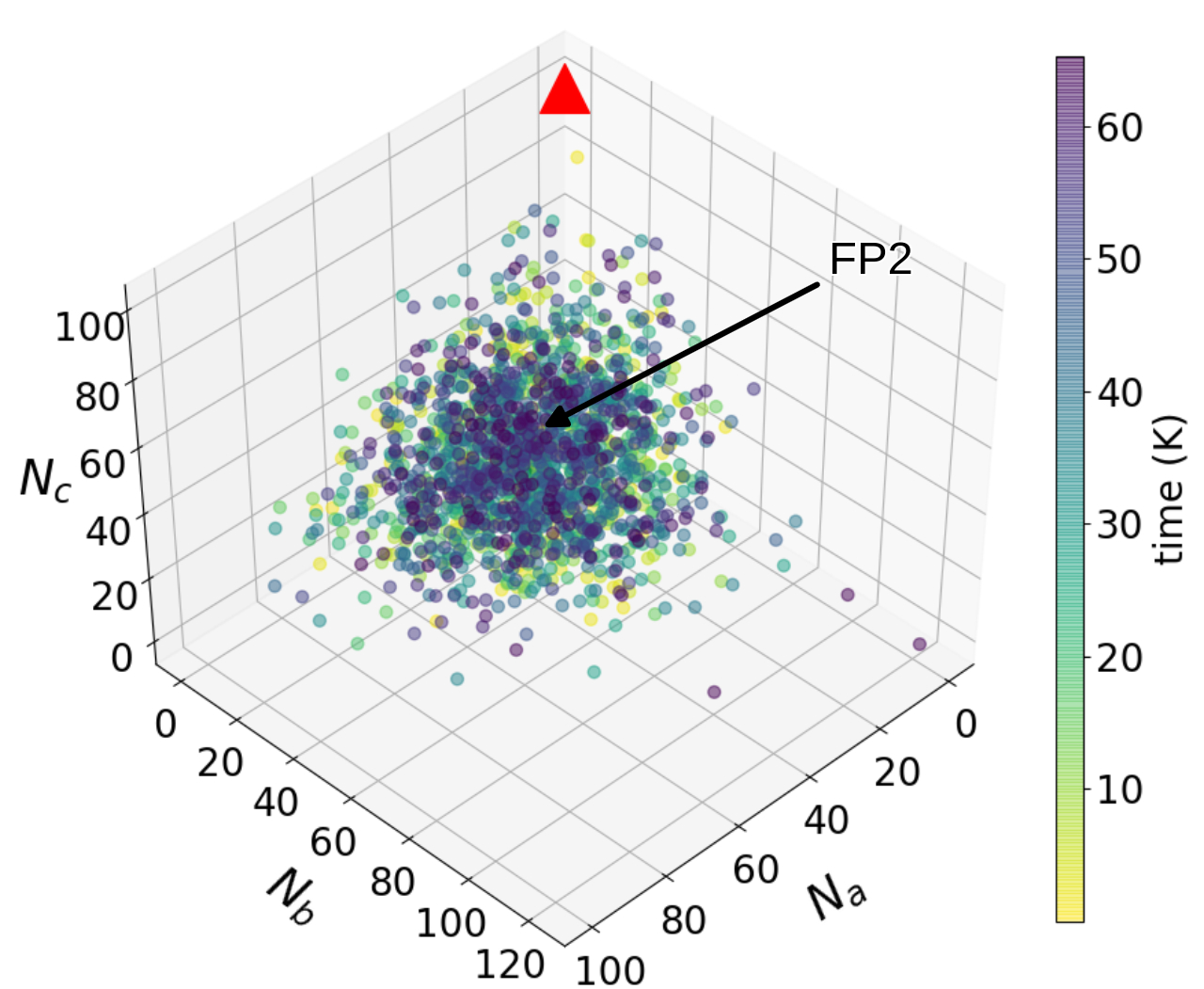}
\endminipage\hfill
\minipage{0.33\textwidth}%
  \includegraphics[width=\linewidth]{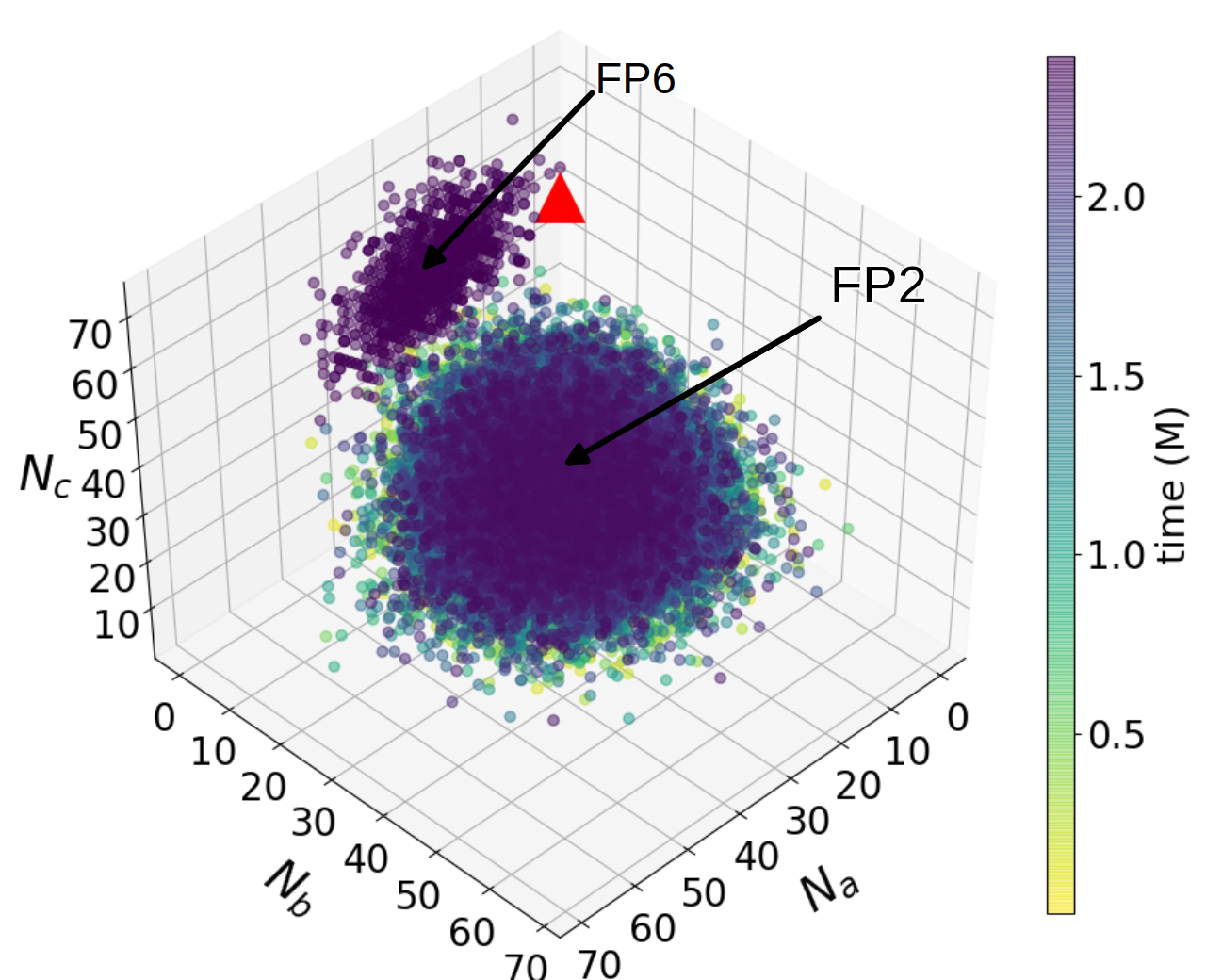}
\endminipage
\caption{Three trajectories of the Gillespie simulations in configuration space from FP$_{2}$ (green triangle) to FP$_{3-5}$ (red triangles) for $\rho=0.1$, $\gamma=1$ in all cases and $\frac{\kappa}{\gamma}=1.8 $ (left)$\ ,1.2$ (center)\, $0.5 $ (right). The color codes the time, light (early), dark (late), measured in the number of Gillespie steps, in units of $10^3$ ((K), center), and $10^6$ ((M), right). The configuration space is spanned by the number $N_\alpha$ of individuals of species $\alpha\in\{a,b,c\}$.}\label{fig:cloudplots}
\end{figure}
at different values of the bifurcation parameter $\frac{\kappa}{\gamma}$ (see figure~\ref{fig:cloudplots}). All three multiplicities of the two-species extinction-fixed points FP$_{3-5}$ are equally likely for a symmetric version of the model. For $\frac{\kappa}{\gamma}$ close to the Hopf bifurcation  (figure~\ref{fig:cloudplots} left), the fluctuations are relatively large and the system approaches FP$_{3-5}$ very quickly as seen in the color map, where color codes time. The system ends up at $(0,0,100)$. For $\frac{\kappa}{\gamma}=1.2$ (figure~\ref{fig:cloudplots} center), the system is  very stable in the deterministic limit and spends most of the time near FP$_{2}$ with coordinates $(46,46,46)$. However, occasional but large fluctuations take the system to FP$_{3-5}$, seen at the corners with two vanishing coordinates, for example $(0,0,100)$. When the system at FP$_{2}$ is far away from the Hopf bifurcation and the fixed point FP$_{3-5}$ is beyond the transcritical bifurcation at  $\frac{\kappa}{\gamma}=0.5$ (figure~\ref{fig:cloudplots} right), the system spends a large portion of time in the vicinity of  FP$_{2}$ with small fluctuations. However, large, but rare fluctuations take the system into the vicinity of single-species extinction-fixed points FP$_{6-8}$ (indicated in the figure is only FP$_6$ with vanishing $N_b$, the number of individuals of type $b$). The system will spend some time in the cloud around FP$_{6-8}$, before yet another large, but rare fluctuation kicks it to FP$_{3-5}$, indicated by the red triangle. As mentioned before, on the time scales, at which Gillespie simulations are run, transitions to FP$1$ are not observed.

\subsection{Distributions of the times to extinction}
\label{RES:Kurtosis}
\begin{figure*}[t!]
  \centering
  \begin{subfigure}[t]{0.32\textwidth}
      \centering
      \includegraphics[height=1.55in]{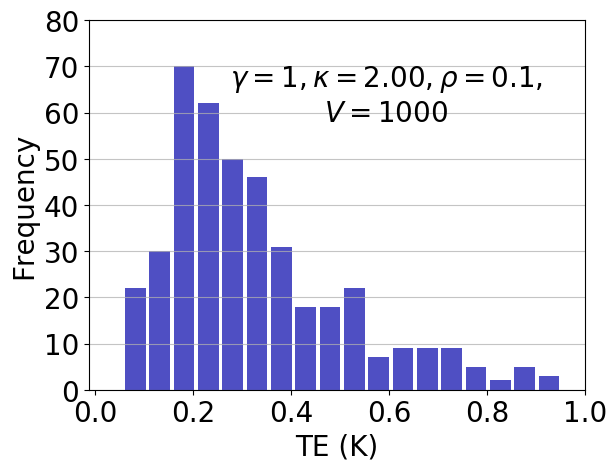}
      \caption{}
  \end{subfigure}%
  \hspace{3mm}
  \begin{subfigure}[t]{0.32\textwidth}
      \centering
      \includegraphics[height=1.55in]{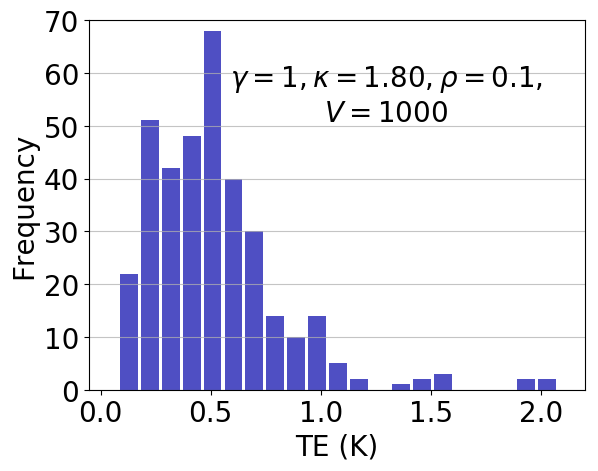}
      \caption{}
  \end{subfigure}
  \begin{subfigure}[t]{0.32\textwidth}
    \centering
    \includegraphics[height=1.55in]{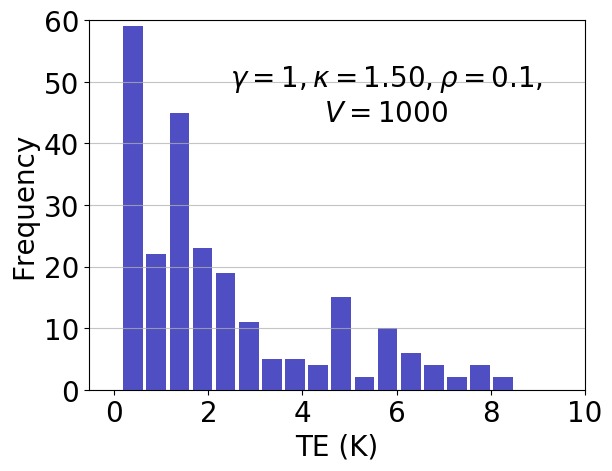}
    \caption{}
\end{subfigure}
  \caption{Probability distribution function of the times to extinction (TE) from FP$_2$ to FP$_{3-5}$ for $\kappa=1.5,\;1.8,\;2.0$ at $\rho=0.1,\ \gamma=1$, V=1000. TEs are measured in units of Gillespie steps.}\label{fig:histograms}
\end{figure*}

So far we have estimated the mean time to extinction that is expected to be quite large as compared to the relaxation time $t_{rel}$ to the stationary state, determined by the deterministic rate equation. Usually,  the MTE should be large if the large fluctuation required by the extinction event is rare. Close to the bifurcation point, large fluctuations are not necessarily rare. Now, it is instructive to measure the  distribution of  times to extinction (TEs) from Gillespie simulations and determine
the histograms, as displayed in figure~\ref{fig:histograms} for different bifurcation parameters $\kappa/\gamma$ with $\gamma=1$ and $\kappa=2$ (left), $\kappa=1.8$ (center) and $\kappa=1.5$ (right). At the bifurcation point, $\kappa/\gamma=2$, large fluctuations are not really rare which lead to extinction. As we know from equation~(\ref{MTE_comp}), the long-time decay of the probability to observe concentrations $n_\alpha>0$ at time $t\gg t_{rel}$ is given by $P(n_\alpha>0,t)=P_{st}e^{-t/\tau}$ with the quasi-stationary distribution $P_{st}(n_\alpha)\propto e^{-s_{st}(n_\alpha) V}=1/\tau$, so for the MTE, estimated as $\tau$, we have MTE$\simeq e^{+s_{st}(n_\alpha) V}$.

\captionsetup[subfigure]{oneside,margin={0.6cm,0cm}}
For short times $t\ll \tau$, the distribution is quasi-stationary, until $t\sim \tau$, when the exponential decay with $t$ sets in.
We note that the parameter choice of figure~\ref{fig:histograms} ($\gamma=1,\rho=0.1,V=1000$) is equivalent to $\gamma=1,\rho=1,V=100$, where  $\gamma=1,\rho=1$ have been used in the WKB-calculations: A change in $\rho$ by a factor $\beta$ can be absorbed in a mere change of the time scale, if at the same time the value of $V$ is scaled with $1/\beta$. The only difference is that for $\rho=0.1,V=1000$  statistics can be gained faster, but in principle we can compare these numerical results with WKB-predictions for the same choice of parameters. The times to extinction are measured in units of thousands of Gillespie steps. Gillespie steps are themselves system-size dependent, as we count the time, at which on average all individuals have been updated once according to the prescribed reactions,  as one Gillespie step. In these units, away from the bifurcation point at $\kappa=1.5$, the longest times to extinction are roughly an order of magnitude larger than at the bifurcation point, the events are more rare. Obviously, at the bifurcation point extinction events at short times occur with a frequency which does not reflect a  probability distribution of survival which decays exponentially with time. Therefore, these events should not be termed rare.

The Gillespie time is given in the number of updating steps of the whole set of individual units. The time in physical units would correspond to the CPU-time in these computer experiments. We cannot verify the precise scaling of the CPU-time with the system size, but once we have to deal with rare events, we may estimate the associated  change in the MTEs if the bifurcation parameter $\kappa$ is varied. A change of $\kappa$ from $2.0$ to $1.5$ leads to a difference in the scaled action from $1.05$ to $1.1$ as we read off from figure~\ref{Action_comparison}. The difference of $0.06$ is a small number, but it gets amplified due to the exponential size dependence of the MTE. The resulting ratio of the MTEs leads to $e^{(1.1-1.05)V}\approx 400$ for $V=100$ (and $\rho=1$). Therefore, the message is that even in the same dynamical regime (for parameters below the Hopf bifurcation), small changes in the bifurcation parameter get amplified by the exponential dependence on the system size and considerably vary the time scale on which rare (extinction) events must be expected.

\section{Conclusions}
\label{RES:Conclusions}
In this work, we have analysed rare event extinction phenomena in the May-Leonard model with cyclic predator-prey interactions. It is seen that ecological diversity in terms of coexistence of all species holds only temporarily: When left alone for long enough, the system will eventually go extinct due to stochastic fluctuations.
The coexistence of interacting species is endangered even when a deterministic description of the system suggests that it is stable.  Usually rare events of large deviations in the population size are thought to be too rare to happen. However, as we have seen, the large deviations although rare can launch a system into extinction of anything from  a single species to all three  species. These extinction phenomena are captured through the WKB-ansatz of the probability distribution of species concentrations, which projects on the rare-events and allows an estimate of the mean-time to extinction.
\par
In contrast to earlier work on single-species extinction, this work studies rare-event extinction phenomena  of several cyclically interacting species.  The WKB-ansatz predicts the extinction along instanton-like trajectories here in a six-dimensional phase space, together with an estimate of the time when the extinction event occurs.
Beyond this estimate of the typical time to extinction (MTE), we analyzed the distribution of the times to extinction as a function of  $\frac{\kappa}{\gamma}$, the ratio of predation to deletion rate that serves as a bifurcation parameter.  Here, this dependence has been studied only for  times to extinction along a specific path (FP$_2$ to FP$_{3-5}$), which was numerically accessible. Interestingly, however, it sheds some light on the sensitivity of the occurrence of rare fluctuations to system parameters. Even within the same model and the same dynamical regime (which allows a stable coexistence of species in the deterministic description), the mean times to extinction change considerably.
\par
Certainly typical for systems with multiple cyclically interacting species is our observation that in combination with symmetries, fixed points have a multiplicity equal to the number of species (here $3$), and beyond this degeneracy further routes to extinction exist in parallel.  We compared the MTE via the scaled action of the direct route to three-species extinction with the indirect route to two-species extinction, followed by the extinction of the only survivor. These times were  similar, but need not agree, as the paths differ. In general, it seems technically challenging for both an analytical and a numerical approach, to  include all possible routes in an  estimate of the MTE. A direct comparison between results for parameter ranges, in which both methods are simultaneously applicable, would be desirable, but in this work the WKB-approach and the Gillespie simulations were feasible in complementary ranges.  Yet, in view of ecological applications it is quite worthwhile to quantify this kind of extinction risk by carefully analyzing the time scales in view of the actual meaning of ``rare". What was termed ``rare" in our context was compatible with times to extinction differing by orders of magnitude.

\section{Acknowledgements}
S.S.'s research was sponsored by the Army Research Office and was accomplished under Grant No. W911NF17-1-0156. The views and conclusions contained in this document are those of the authors and should not be interpreted as representing the official policies, either expressed or implied, of the Army Research Office or the U.S. Government. The U.S. Government is authorized to reproduce and distribute reprints for Government purposes notwithstanding any copyright notation herein. We are also grateful to the German Research Foundation (DFG, grant number ME-1332/27-1) for financial support of mutual visits at Jacobs University Bremen and Virginia Tech Blacksburg University in 2016 and 2017, where part of this work was initiated. We also thank Jason Hindes, Darka Labavi\'{c}, Uwe C. T\"auber, Priyanka, Shengfeng Deng for fruitful discussions and comments regarding this work.
\medskip
\section*{References}
\bibliographystyle{iopart-num}
\bibliography{revised_hmortmanns}
\end{document}